\documentclass[11pt]{article}
\usepackage{amsfonts,amsbsy,bm,euscript,mathrsfs}
\usepackage{amssymb,stmaryrd,faktor}
\usepackage[tbtags]{amsmath}
\usepackage[title,titletoc]{appendix}
\usepackage[bookmarks=true,colorlinks=true,linkcolor=blue,citecolor=blue,urlcolor=blue,bookmarksnumbered]{hyperref}
\usepackage{graphics}
\usepackage{tikz}
\usetikzlibrary{matrix,arrows}

\newcommand{\mathsym}[1]{{}}
\makeatletter
\renewcommand\section{\@startsection {section}{1}{\z@}
{-3.5ex \@plus -1ex \@minus -.2ex}
{2.3ex \@plus.2ex}
{\normalfont\large\bfseries}}
\renewcommand\subsection{\@startsection{subsection}{2}{\z@}
{-3.25ex\@plus -1ex \@minus -.2ex}
{1.5ex \@plus.2ex}
{\normalfont\normalsize\bfseries}}
\renewcommand\subsubsection{\@startsection{subsubsection}{2}{\z@}
{-3.25ex\@plus -1ex \@minus -.2ex}
{1.5ex \@plus.2ex}
{\normalfont\small\bfseries}}
\makeatother
\newcommand{\mc}{\mathcal }

\begin{document}

\overfullrule=0pt
\parskip=2pt
\parindent=12pt
\headheight=0in \headsep=0in \topmargin=0in \oddsidemargin=0in

\vspace{ -3cm}
\thispagestyle{empty}
\vspace{-1cm}

\rightline{ Imperial-TP-AT-2014-06}

\begin{center}
\vspace{1cm}
{\Large\bf   Vectorial  AdS$_5$/CFT$_4$ duality  for spin-one  boundary  theory    }
\vspace{1.5cm}

{{\large Matteo Beccaria}$^{a,}$\footnote{matteo.beccaria@le.infn.it} and {\large Arkady A. Tseytlin}$^{b,}$\footnote{Also at Lebedev Institute, Moscow. tseytlin@imperial.ac.uk }}\\

\vskip 0.6cm

{\em $^{a}$    Dipartimento di Matematica e Fisica Ennio De Giorgi,\\
Universit\`a del Salento \& INFN, Via Arnesano, 73100 Lecce, 
Italy }

\vskip 0.3cm

{\em
$^{b}$ The Blackett Laboratory, Imperial College,
London SW7 2AZ, U.K.
}

\vspace{.2cm}

\end{center}

\begin{abstract}
\noindent
We consider an  example  of   vectorial AdS$_5$/CFT$_4$ duality when the boundary theory is described  
 by   free  $N$ complex or real Maxwell fields.  It is 
   dual to a particular (``type C'')  higher spin theory in AdS$_5$ containing fields in  special mixed-symmetry representations. 
 We  extend   the  study   of this  theory 
   in arXiv:1410.3273 by  deriving the expression  for the large $N$  limit  of the 
 corresponding  singlet-sector partition function on $S^1 \times S^3$. 
 We find that  in both complex $U(N)$ and real $O(N)$ invariant cases   the form of the one-particle partition function  is as  required by the AdS/CFT duality. We also  discuss  the  matching of the Casimir energy on $S^3$  by assuming an integer   shift  in  the bulk theory coupling.
\end{abstract}

\newpage

\setcounter{equation}{0}
\setcounter{footnote}{0}
\setcounter{section}{0}

\newcommand{\be}{\begin{equation}}
\newcommand{\ee}{\end{equation}}

\newcommand{\wt}{\widetilde}
\newcommand{\wh}{\widehat}
\newcommand{\mk}{\mathfrak}
\newcommand{\hyper}[3]{{}_{3}F_{2}\left(\left.\begin{array}{c} #1 \\ #2 \end{array}\right | #3\right)}
\newcommand{\hs}{\mbox{hs}[\lambda]}
\newcommand{\sref}[1]{Sec.~(\ref{#1})}
\newcommand{\mycheck}[1]{\footnote{\red{\tt #1}}}
\newcommand{\figref}[1]{Fig.~(\ref{fig:#1})}

\newcommand{\la}{\longrightarrow}
\newcommand{\irrep}[3]{(\mathbf{#1},\mathbf{#2})_{#3}}

\newcommand{\ds}{\displaystyle}
\newcommand{\vD}{\vec{\slashed{\DD}}}
\newcommand{\DD}{{\mathcal D}}
    \renewcommand{\(}{\left(}
    \renewcommand{\)}{\right)}
    \newcommand{\eq}[1]{(\ref{#1})}
    \newcommand{\ph}{{\rm ph}}
    \newcommand{\mir}{{\rm mir}}
    \newcommand{\beq}{\begin{equation}}
    \newcommand{\eeq}{\end{equation}}
    \newcommand\bea{\begin{eqnarray}}
    \newcommand\eea{\end{eqnarray}}
        \renewcommand{\d}{\partial}

\newcommand{\CC}{{\mathcal C}} 
\newcommand{\sql}{{ \sqrt\lambda}}

\def \Z {{\cal Z}}\def \A {{\cal A}}\def \N {{\cal N}}
\def \del{ \partial}
\def \la {\label}
\newcommand{\rf}[1]{(\ref{#1})}
\def\ov{\over}
\def\no{\nonumber} \def \aa {{\rm a}}

\def \ci {\cite}
\def \p {\phi}
\def \m {\mu}\def \n {\nu} 
\def \ed {\end{document}}
\def \l {\lambda} \def \r {\rho} 
\def \cP {{\cal P}}
\def \foot {\footnote}
\def \b {\beta} 
\def \dd {{\rm d}} 
\def \om {\omega} 
\def \tr {{\rm tr}}
\def \D {\Delta} 
\def \vp {\varphi} 
\def \mR {{\mathbb  R}} \def \ha {{{1 \ov 2}}}
\def \bD {{\mathbf{D}}}
\def \wtd {\widetilde}
\def \cc {{\rm c}}
\def \aa {{\rm a} }
\def \z {\zeta} 

\def \De {\Delta} 
\def \ads {AdS$_{5}$\ }
\def \te {\textstyle} \def \iffa {\iffalse} 

\def \ha {{\te {1 \ov 2}}}

 \def  \ba { \begin{align} }
 \def  \ea { \end{align} }

\def \gg   {{\rm g}}
\def \cc    {{\rm c}} 
\def \aa  {{\rm a}}

\def \ep {\epsilon}
 \def \k {\kappa} \def \r {\rho} 

\def \RR {{\rm R}}
\def \OO {{\cal O}} 
\def \edd {\end{document}} 
\def \td {\tilde} 

\def \tO {{\td \OO}}\def \rZ {{\rm Z}}

 \def \Deltat  {{\bar \OO}}
 
\def \de {\delta} 
\def \KK {{\rm K}}

\def \zz {\Z}

\def \P {\Phi}
\def \te {\textstyle}
\def \ha {{\te { 1 \ov 2}}}
\def \for {{\te { 1 \ov 4}} }
\def \edo {\end{document}}
\def \dd {{\rm d}} 
\def \ff {{\cal F}}
\def \AA   {{\cal A}}
\newcommand{\enn} {\end{align}}
\def \FF {{\cal F}} 
\def \a {\alpha}\def \fo {{\te {1 \ov 4}}}
\def \G {\Gamma}  
 \def \wZ {{ \mZ}}
\def \cD {{\cal D}}

\def \mZ  {{\rm Z}}
  \def \rk {{\rm k}}
\newcommand{\ts}{\textstyle}

\def \OO  {{\mc O}}\def \Tr {{\rm Tr }}   \def \tr  {{\rm tr }}   

     \def \adss {AdS$_5$.\ }

\renewcommand{\theequation}{1.\arabic{equation}}
\setcounter{equation}{0}
\section{Introduction} 
%
In addition to ``adjoint'' AdS/CFT duality    between gauge   theory at the boundary  and string theory in the bulk 
there is a simpler example of ``vectorial'' AdS/CFT relating  singlet sector  of 
  conserved currents of a  CFT  described  by  $N$ free  massless   fields  
 in a   vector  representation of $U(N)$  or $O(N)$  to  a  Vasiliev-type theory in AdS. 
   This  duality is not restricted to 
 original examples in  $d=3$ \cite{Klebanov:2002ja,Sezgin:2003pt}:    
  generalizations to  $d>3$  
   were   studied, e.g.,   in \ci{Didenko:2012vh,
   Bekaert:2013zya,Giombi:2013yva,Giombi:2013fka, Giombi:2014iua,Giombi:2014yra}.\foot{An early discussion of $d=4$ 
   case based on gauged $O(N)$ model appeared in \ci{Schnitzer:2003zr}.}
While in   $d=3$  the  only option to get a   unitary theory  with a higher spin symmetry is 
 to use   free scalars or  spin $\ha$  fermions  
     \ci{Maldacena:2011jn} in the $d=4$  case  that  we will be interested in here
  one is also allowed to consider free spin 1  fields \ci{Boulanger:2013zza,Stanev:2013qra,Alba:2013yda}.\foot{Higher-spin 
  symmetries of  such 
  boundary models   and  their higher-spin and  super-extensions where found in \ci{vasiliev:2001}.
  More generally,
one may also  attempt to  use   a higher-spin  doubletons at  the boundary; this possibility was noticed
in \ci{Bekaert:2009fg}   where the corresponding  higher spin  algebras were studied.}

 The   corresponding conserved currents that  appear    in the  product of two   spin $1$   doubletons 
   (as in \ci{Gunaydin:1998sw,Ferrara:1998jm,
   Sezgin:2001zs,Vasiliev:2004cm,Dolan:2005wy, Boulanger:2011se})
     are  in  specific 
      mixed-symmetry representations of $SO(2,4)$.\foot{Mixed-symmetry fields  in \ads  and the associated currents  were  discussed, e.g.,    in 
    \ci{Metsaev:1995re,Alkalaev:2003qv, Alkalaev:2012rg}.}
   The singlet sector of a theory of $N$ complex  or real Maxwell  vectors 
   invariant under $U(N)$ or $O(N)$    should then   be dual to a particular   version of 
    higher spin theory in \ads involving mixed-symmetry fields. 
    This theory was    called  ``type C''   
      in   \ci{Beccaria:2014xda} by analogy with type A  theory  dual to $N$  boundary 
     scalars  and type B theory dual to   $N$ spin $\ha$ boundary fermions. 
    Here we will   elaborate on  its  discussion  in  \ci{Beccaria:2014xda}  by 
     directly computing  the   singlet-sector partition in this theory  
     and explaining the matching of the \ads  vacuum energy 
     in the bulk and the  $S^3$ Casimir energy at the boundary.

\renewcommand{\theequation}{2.\arabic{equation}}
\setcounter{equation}{0}
    \section{Representation  content and relations   between characters}
    
    Let us first   review  the case when the boundary theory  is  
    described by $N$ complex  or real  scalars. We shall use the notation $(\De; j_1,j_2)$ 
    for generic ``massive'' representation of $SO(2,4)$, with the ``massless'' case   corresponding to
     $\De=2 + j_1 + j_2$  with $j_1 j_2 >0$.\foot{Here 
    $(j_1,j_2)$ are $SU(2) \times SU(2)$  weights (with total spin  being  $s=j_1+j_2$)  and we shall use the notation 
     $(\De;j_1, j_2)_c= (\De;j_1, j_2) + (\De;j_2, j_1)$.  } 
  Also, $  \{j, 0\}$ and $  \{0,j\} $  with $\De= 1 + j$   will denote  spin $j$   doubleton representation.  The  spectrum of states 
  in ``non-minimal''  type A theory dual to  complex $U(N)$ scalar theory   may be 
   found  from the decomposition of the product of two $j=0$ doubletons \ci{Vasiliev:2004cm} (generalizing the $d=3$ relation of  \ci{Flato:1978qz})
\be
{\rm non-minimal \  A}: \ \  \qquad  \   \{0,0)\otimes \,\{0,0\} = \mc (2;\,0,0)+\bigoplus_{s=1}^{\infty}\mc (2+s; \ts\frac{s}{2},\frac{s}{2}) \ .   \la{a5}
\ee
    Here  the representations 
    $ (2+s; \ts\frac{s}{2},\frac{s}{2})$ correspond to  conserved currents dual to 
     massless totally symmetric  spin $s$ fields in AdS$_5$. 
      In the ``minimal'' type A theory case   dual to the  real scalar  $O(N)$ theory 
     one is to project out all odd spin fields (the corresponding currents  become trivial), i.e. to ``symmetrize''  the product 
   \be
{\rm minimal\  A}: \ \  \qquad  \  \big[ \{0,0) \otimes \,\{0,0\}\big]_{\rm sym} = 
\mc (2;\,0,0)+\bigoplus_{s=2,4,...}^{\infty}\mc (2+s; \ts\frac{s}{2},\frac{s}{2}) \ .   \la{a6}
\ee  
 Let us define the  (``blind'') characters for the  basic representations  ($\De_0= 2 + j_1 + j_2$) \ci{Dolan:2005wy}\foot{We follow the notation of \ci{Beccaria:2014xda}. Eq.\rf{a8}  applies   also to the case of massive  selfdual  representations  that appear when $j_1 j_2=0$ 
and $\De > 1 + j_1 + j_2$.}
\begin{align}
{\rm  ``massive"}:&  \ \ \  \mZ(\Delta; j_{1}, j_{2}) =\frac{q^{\Delta}}{(1-q)^{4}} (2j_{1}+1)\,(2j_{2}+1)\ ,\la{a8} \\
{\rm ``massless"}:& \ \ \    \mZ(\De_0 ; j_{1}, j_{2}) = \frac{q^{\De_0}}{(1-q)^{4}}\,\big[(2j_{1}+1)(2j_{2}+1)-4\,q\,j_{1}\,j_{2}\big]\ ,\la{a9} \\
{\rm ``doubleton"}:& \ \ \    \mZ ( \{j, 0\}) = \mZ(\{0, j\}) = \frac{q^{1 + j}}{(1-q)^{3}}\big[2j+1 -q\,(2j-1)\big]\ . \la{a10}
\end{align}
Here   \rf{a8}/\rf{a9}  has the interpretation of   one-particle partition function  $ \Z(\Delta_0; j_{1}, j_{2})$ 
   for the  corresponding  massive/massless  5d field
   in \ads  with standard  boundary conditions. 
  The character identities   which are the counterparts   of \rf{a5} and \rf{a6} are \ci{Dolan:2005wy}
   \be
   {\rm non-minimal \  A}: \ \    \  \big[\mZ  ( \{0, 0\})\big] ^{2}= \mZ(2; 0, 0)+
\sum_{s=1}^{\infty} \mZ(2+s; \ts\frac{s}{2}, \frac{s}{2})\  , \qquad \qquad 
\la{a166}
\ee
and \ci{Giombi:2014yra}
  \be 
 {\rm minimal \  A}: \ \  \ha    \big[\mZ  ( \{0, 0\})\big] ^{2}+ \ha \, \big[ \mZ  ( \{0, 0\})\big]_{q\to  q^2} = \mZ(2; 0, 0)+
\sum_{s=2, 4, \dots}^{\infty} \mZ(2+s; \ts\frac{s}{2}, \frac{s}{2}) \ . 
\la{a16}
\ee
The validity  of \rf{a16}   can be verified  directly, but  the  group-theoretic origin of the  combination in the l.h.s.  is  not 
 obvious. The l.h.s. parts   of \rf{a166} and \rf{a16}   have the interpretation   of the  large $N$ limit of 
  singlet-sector one-particle 
partition functions 
of the $U(N)$   \ci{Shenker:2011zf} and the $O(N)$  \ci{Giombi:2014yra} scalar  theories. The r.h.s. parts 
are interpreted as the  total  contribution to the one-particle partition function 
  of  corresponding higher spin theory defined on thermal quotient  of \ads \ci{Gopakumar:2011qs,Gupta:2012he}.

The same   construction can  be repeated for spin $\ha$  boundary theory when the role  of $j=0$ doubleton is
 played by the $j=\ha$  one,  leading to the spectrum of type B   higher spin theory in \ads  \ci{Giombi:2014yra,Beccaria:2014xda}.
Similarly,  taking the product of two  $j=1$ doubletons  $\{1,0\}_c= \{1,0\} + \{0,1\}$  (that may be associated with the 
self-dual and anti self-dual   parts of Maxwell field  strength $F_{mn}$) we get the spectrum of  
conserved currents   and other primary fields    and thus  the content 
of the   corresponding type C theory in \ads  \ci{Beccaria:2014xda}.
In general, the products of two doubleton representations decompose as follows   \ci{Vasiliev:2004cm,Dolan:2005wy}
\begin{align}
\{j, 0\}\otimes \,\{j', 0\} &= \bigoplus_{k=|j-j'|}^{j+j'} (2+j+j';\,k ,  0)+\bigoplus_{k=1}^{\infty}(2 + j + j' + k;  j+j'+\ts\frac{k}{2}, \frac{k}{2}) \ ,\la{a2} \\
\{0,j\}\otimes \,\{0,j'\} &= \bigoplus_{k=|j-j'|}^{j+j'}  (2+j+j';\, 0,k)+\bigoplus_{k=1}^{\infty}(2 + j + j' + k; \ts \frac{k}{2}, j+j'+\ts\frac{k}{2}) \ ,\la{a3} \\
\{j, 0\}\otimes \{0, j'\} &= \bigoplus_{k=0}^{\infty}(2+ j + j' + k;  j+\ts\frac{k}{2}, j'+\frac{k}{2})\ . \la{a4}
\end{align}
For an ``unsymmetrized''
   product  of  two spin 1  doubletons one then  finds the spectrum of the non-minimal type C theory 
dual to   the singlet sector of $N$ complex Maxwell   fields   at the boundary
\begin{align}
& {\rm non-minimal \ C}: \qquad  \qquad  \{1, 0\}_c  \otimes    \{1, 0\}_c\equiv  
\big(  \{1, 0\}  +  \{0, 1\}   \big)  \otimes   \big(  \{1, 0\}  +  \{0, 1\}   \big)  \no \\
&\qquad =2\,(4;\,0,0)+(4;\,1,0)_{c}+  (4;\,2,0)_{c}  + {
2\,\bigoplus_{k=0}^{\infty}(4+k; \ts\frac{k+2}{2},\frac{k+2}{2}) }+
\bigoplus_{k=1}^{\infty}(4+k; {\ts 2+\frac{k}{2},\frac{k}{2}})_{c}\no  \\
&\qquad = 2\,(4;\,0,0)+(4;\,1,0)_{c}+
2\,\bigoplus_{s=2}^{\infty}(2+s;  {\ts\frac{s}{2},\frac{s}{2})} +\bigoplus_{s=2}^{\infty}(2+s; {\ts\frac{s+2}{2},\frac{s-2}{2}})_{c}\ . \la{a7}  \end{align}
 In addition to two  infinite series of massless  spin $s \ge 2$  fields  it contains also a massive scalar and pseudoscalar in 
 representation $(4; 0, 0)$
 (dual to $F^*_{mn} F^{mn}$ and $F^*_{mn} \td F^{mn}$)
 and the  rank 2   antisymmetric   tensor  in self-dual and anti self-dual massive representaion $(4; 1,0)_c$
  (dual to  $F^*_{m[n} F_{k]m}$).
  The   spectrum of minimal type C theory dual to  real $O(N)$  Maxwell theory is found  by 
  projecting out   one parity-odd  symmetric tensor states and   odd-spin  mixed-symmetry states: 
\begin{align}
& {\rm minimal \ C}: \ \ \qquad   \big[\{1, 0\}_c  \otimes    \{1, 0\}_c\big]_{\rm sym}  \no \\
&  \qquad  \qquad \qquad  \qquad \qquad = 2\,(4;\,0,0) +
\bigoplus_{s=2}^{\infty}(2+s;  {\ts\frac{s}{2},\frac{s}{2})} +\bigoplus_{s=2,4,...}^{\infty}(2+s; {\ts\frac{s+2}{2},\frac{s-2}{2}})_{c}\ . \la{a89}  \end{align}
The corresponding character relations are  counterparts of \rf{a166}  and \rf{a16}  in type A case:
\ba
\no 
 &{\rm non-minimal \ C}: \ \ \qquad    \big[ \mZ  ( \{1, 0\}_c)\big] ^{2} \no\\
 &\qquad \quad \   =  2\, \wZ(4; 0, 0)+   \wZ(4; 1,0)_{c} +    2  \sum_{s=2}^{\infty}\mZ (2+s; {\ts\frac{s}{2}, \frac{s}{2}})  + 
  \sum_{s=2}^{\infty} 
\mZ(2+s; {\ts\frac{s+2}{2}, \frac{s-2}{2}})_{c} \ ,  \la{a20}   \\
& {\rm minimal \ C}: \ \ \qquad\qquad  \ha    \big[\mZ  ( \{1, 0\}_c)\big] ^{2}+ \ha  \big[\mZ  ( \{1, 0\}_c)\big]_{q\to  q^2} \no \\
  &\qquad \qquad    =  \  2\, \wZ(4; 0, 0)+\sum_{s=2}^{\infty}\mZ (2+s; {\ts\frac{s}{2}, \frac{s}{2}})   
   +  \sum_{s=2,4,\dots}^{\infty} \mZ(2+s; {\ts\frac{s+2}{2}, \frac{s-2}{2}})_{c}  \ . \la{a21} \end{align}
Like in  the  type A case and the  type B cases \ci{Giombi:2014yra}
these relations between   characters have again a field theory  or AdS/CFT interpretation. 
As we shall show   below,  the l.h.s. of \rf{a20}  is  the one-particle 
partition function  representing the leading term in the  large $N$  limit  of the singlet-sector  partition function 
of the theory of $N$ complex Maxwell fields, while the l.h.s. of \rf{a21} corresponds to the real $O(N)$ Maxwell theory case.

\renewcommand{\theequation}{3.\arabic{equation}}
\setcounter{equation}{0}
\section{Singlet-sector partition function in  the theory of $N$  Maxwell fields} 

Let us first recall the  expression for the partition function of   one scalar  field on $S^1 \times S^3$, 
then  consider $N$ scalars  and 
impose the  singlet-sector 
 constraint and,    finally,   generalize  to  the  case of Maxwell vectors  instead of scalars.

For a conformal  scalar   we get  the partition function (we assume  $S^3$ to have  unit radius and length of $S^1$ to be $\b$) 
\be \la{25} 
 Z_{0}  = (\det\, \OO_0)^{-1/2} \ , \ \ \ \ \ \ \
 \OO_0= - D^2 + {1\ov 6} R=   -\partial_{0}^{2}-\mathbf{D}^{2}+  1 \ .  
\ee
Using the  eigenvalues  of the Laplacian $-\mathbf{D}^{2}$  on $S^{3}$ 
we get the  eigenvalues  and multiplicities   of  $\OO_0$  ($ k=0, \pm 1, ..., \ \ n = 0, 1, 2, \dots$)
\be
\lambda_{k,n} = w^2_k +\omega_{n}^{2}, \qquad   w_k= {2\pi k\beta^{-1}} \ , \quad 
\omega_{n} = n+ 1 \ , \qquad \dd_n = (n+1) (n+2) \  .\la{29} 
\ee
Thus   $\log  Z_{0}= -\ha \log \det \OO_0= -  \ha \sum_{k,n} \dd_n \log \lambda_{k,n} $. 

In general,     for the  bosonic partition function  we have  ($ q\equiv e^{-\beta} $)
\ba  &\qquad  \G= -\log Z =  \b E_c + F(\b)  \ , \ \ \ \ \ \qquad       E_c = \ha \sum_n \dd_n \ \om_n \ , \la{101} \\
 & F =  \sum_n {\dd}_n \log (  1 - e^{-\beta \omega_n})  = -\sum_{m=1}^\infty  \frac{1}{m}\,  \Z(q^m)\ , \ \ \ \qquad  \ \ 
   \Z(q) =  \sum_n  {\rm d}_n\, e^{-\beta\,\omega_n} 
 \ , \la{2222}
\end{align}
where  $E_c$ is Casimir energy on $S^3$     and $\Z$ is one-particle partition function.  
For
the massless  4d scalar  $\Z$ thus   has     the same expression as  the character for  spin 0 doubleton in \rf{a10}
\be
\Z_{0} =  \mZ(\{ 0,0\}) = \sum_{n=0}^{\infty} (n+1) (n+2) \,  q^{ n+1} = \frac{q (1+q)}{(1-q)^{3}}\ .  
\la{210} \ee
Next,   let us    consider the  singlet partition function for $N$ complex  scalars in $d=4$ 
(see    \ci{Shenker:2011zf,Giombi:2014yra}  and  also 
 \ci{Sundborg:1999ue,Aharony:2003sx, Schnitzer:2006xz} for relevant earlier work). 
One way to define  it is to  gauge the  $U(N)$ symmetry and consider the coupling of $N$   scalars $\P_r$ to $U(N)$ gauge field
$\AA_m$ with strength $\FF_{mn}$
\be 
L=  D_m \P^*_r\,  D^m \P_r     +   { 1 \ov 4 g^2}  \tr \, \FF_{mn} \FF^{mn}  \ . \la{71} \ee
We shall understand  the  limit $g \to 0$  as restricting the  path integral to pure-gauge fields $\AA_m$.  
In the case  of $S^1 \times S^3$  there is a non-trivial holonomy $U$ of $\AA_0$ 
along $S^1$  that cannot be gauged away.  The remaining path integral will then be 
over $\P_r$ and $U$   or  phases $\a_r$ in 
  \be 
   \AA_0 = U^{-1} \del_0 U, \ \  \ \ \ U= {\rm diag }(e^{i\a_1\tau}, ..., e^{i\a_N\tau}) , \ \ \ \ \ 
  \tau = \b^{-1}  x_0\in (0,1)\ . \la{177} \ee
  We thus  get the  following expression for  the singlet partition function of $U(N)$ scalars 
  \ba 
&\qquad \qquad \qquad  \hat Z  = \int  \prod^N_{r=1}  d\a_r  \ e^{ -  \G (\a;\b)} \ , \la{55}     \\
&   
   \G (\a;\b)  = \m (\a)  + \bar F (\a; \b)  \ , \ \ \ \ \ \ \ \ \qquad 
\m (\a)  = -\ha  \sum^N_{r\not= s=1} \ln \sin^2{\te  { \a_r - \a_s \ov  2}}   \ ,\la{5b} \\
 &\bar F = \ln \det  [ -  (\del_0  + \AA_0 )^2  -  \bD^2    + 1 ] \ = 
  \sum_{r=1}^N \sum_{k=-\infty}^\infty \sum_{n=0}^\infty \dd_n \, 
  \ln \Big[ { ({2\pi k+ \a_r })^2  } \b^{-2}    + \omega_n^2  \Big]\ .  
 \la{5a}
\end{align} 
Here $\m(\a)$ is the contribution of the $U(N)$ invariant 
measure $[U^{-1} d U]$  and 
$\dd_n$    and $\omega_n$ are as in \rf{29},\rf{210}. 
To take the  large $N$ limit  let us introduce the normalized  ($\int_{-\pi}^{\pi} d\alpha\, \rho(\alpha)=1$) 
 eigenvalue density $\r(\a)$   
 and  replace  the integral
over $\a_r$ by the path integral over  the periodic  field  $\r(\a)$ defined  on a circle 
\ba 
 &\qquad \qquad  \hat Z\Big|_{N \to \infty}   =  \int  [d\r]   \ e^{ -  {\rm F}  (\r;\b)} \  , \la{711}\\
 & \la{72}
  {\rm F}  =    N^2 \int d\a d \a'\   K(\a- \a') \, \r(\a)\,  \r(\a')    +   2 N  \int d \a \, \r(\a)\   Q (\a; \b) \ , \\
& \la{73}
K(\a) = -   \ha   \ln  \big(2 - 2 \cos\a\big) 
 , \ \ \  \ \  Q(\a; \b)    = \ha    \sum_{k=-\infty}^\infty \sum_{n=0}^\infty \dd_n  \ln \Big[ { ({2\pi k+ \a })^2 }  \b^{-2}   + \om_n^2  \Big].
\end{align}
Here the $N^2$ term came from the measure term $\mu$ in \rf{5b}.
Isolating the Casimir energy part 
 and rearranging the sum  we get 
\ba
& \la{9}  Q (\a; \b)= \b E_c   + \bar Q   \ , \ \ \ \ \   E_c=  \ha \sum_{n=0}^\infty \dd_n \ \omega_n  \ , 
\ \ \ \   \bar Q (\a; \b)   =   \sum_{m=1}^\infty  c_m(\b) \,     \cos (m\a)  \ , \\ 
& \la{10} 
   c_m (\b) = - { 1 \ov m} \zz_0 (m \b) \ ,   \ \ \qquad \qquad   \zz_0 ( \b)  =       \sum_{n=0}^\infty   \dd_n \, e^{-\b \om_n} \ .   \end{align}
 Splitting  the normalized periodic function $\r(\a)$ into   the constant  and non-constant  parts,  
 $  \rho(\alpha)=\frac{1}{2\pi}+
 \td \rho(\alpha) $,  we can write 
  \rf{72} as   (using  that $\bar Q$ in \rf{9}  does not couple to the constant part of $\r$) 
 \be 
 \la{14}
  {\rm F}  = 2 N \b E_c +    N^2  \int d\a d \a'\,  K(\a- \a') \, \td \r(\a) \, \td \r(\a')    + 
    2 N \int d \a \, \td \r(\a)\,  \bar Q (\a;  \b)\ .
\ee
Integrating over  $\td \r$ (or, equivalently,  evaluating the path integral  at the large $N$ saddle point)  gives finally 
\ba &\qquad \qquad 
\hat \G =  -\log \hat Z \Big|_{N \to \infty}   =2 N \b E_c  +    \hat  F (\b )  \ , \ \ \ \ \ \la{178} \\  &
 \hat  F =   -    \sum_{m=1}^\infty  m  \big[c_m(\b)\big]^2
 =  - \sum_{m=1}^\infty   { 1 \ov m}  \hat \Z( m\b)   \ , \ \ \ \ \ \ \ \ \ \ \   
  \hat \Z
   (\b) =  \big[ \zz_0 (\b)\big]^2     \  .  \la{17a}
\end{align} 
Thus the    one-particle  partition function corresponding to   the large $N$ limit of the singlet-sector 
  partition function 
 is given by the {square}  of the  free scalar one in \rf{210} \ci{Shenker:2011zf}. 
 Repeating this argument  in the real $O(N)$ scalar case  one finds  a similar  result
 \ci{Giombi:2014yra}, i.e.
 \ba   &
  \hat \Z_{\rm U(N)} 
   (\b) =   \big[ \zz_0 (\b)\big]^2   \  ,   \la{17} \\
  &
  \hat \Z_{\rm O(N)}   (\b) =  \ha \big[ \zz_0 (\b)\big]^2   + \ha \Z_0 (2\b)    \  .  \la{777}
\end{align} 
These are exactly the   expressions that appear in the l.h.s. parts of the 
character relations \rf{a166} and \rf{a16}.

The leading Casimir energy term in \rf{178} is the same as in \rf{101} for $2N$ real scalars, 
while the  non-trivial $\b$-dependent part $ \hat  F (\b )$   is  of order $N^0$  rather than of  order $N$ 
  when the singlet constraint is not imposed. 

While  the  scalar one-particle partition function $\Z_0$  \rf{10}   counts all 
 operators built out of one scalar  and its derivatives modulo equations of motion,  
 the  singlet partition function $\hat \Z$ counts   all bilinear   spin $s$ currents 
  modulo the conservation condition. Equivalently, it is the  total one-particle partition function for  the theory of massless 
  higher spin  fields in  thermal cover of \ads  \ci{Shenker:2011zf,Giombi:2014yra}.
 Similar results are found  \ci{Giombi:2014yra} in the spin $\ha$ case  with $\Z_0 \to \Z_{1\ov 2}  = { 4 q^{3\ov 2} \ov (1-q)^3}$
 and the sign plus  in  \rf{177} replaced by the sign  minus.

Let us  now  repeat the same  computation using  the 
  Maxwell   action $S_1= \frac{1}{4}\int d^4 x \sqrt g \,  F_{\mu\nu}\,F^{\mu\nu}$ instead of the scalar one. 
 For   one real Maxwell field 
   we get  the  well-known expression  for the  curved-space partition function  in the  usual covariant Feynman gauge   
\be
Z_1 = \frac{\det(-D^{2})}{\left[\det(- g_{\mu\nu} D^{2}+R_{\mu\nu})\right]^{1/2}}\ , \la{211} 
\ee
Specializing to the $S^1 \times S^3$  case   (splitting  
    $A_{\mu} = (A_{0}, A_{i})$  \ $i,j=1,2,3$) 
we find   \ci{Beccaria:2014jxa}
\be\la{212}
Z_1 = \left[\frac{\det(-D^{2})}{\det(-g_{ij} D^{2}+ R_{ij})}\right]^{1/2}  = 
 \frac{1}{\big[\det(-g_{ij} D^{2}+  R_{ij} )_{\perp}\big]^{1/2}}
=  \frac{1}{\big[\det \, \OO_{1\, \perp}\big]^{1/2}} \ ,
\ee
where  $\OO_{1\, ij} = (- \del_0^2  - \bD^2 + 2 )_{ij} $,   and  we 
split the 3-vector field operator  into the  transverse ($D^{i}A_{i,\perp}=0$)  and longitudinal parts.
The same   expression   can be obtained  directly 
by choosing the $D^iA_i=0$  gauge  in the  path integral,  where the action becomes   \be L_1 =  {1 \ov 2}  \del_0 A_i \del_0 A^i  +  {{1 \ov 4}} F_{ij} F^{ij}  +    {1 \ov 2}  D^i A_0  D_i A_0 \ . 
   \la{a0} \ee 
The contribution  of the  integral over $A_0$ cancels against the 
ghost  or measure determinant giving back \rf{212}. 

From the spectrum  of transverse 3-vector  Laplacian $(-\bD^{2})_{1\,  \perp}$   on $S^3$   
 the spectrum   of $\OO_{1\, \perp}$ in \rf{212}  is found to be 
  is 
\be
\lambda_{k,n} = w^2_k + \omega_n^2 \ , \qquad 
\  \omega_n= n+2\ ,\qquad  \dd_{n} = 2(n+1)(n+3) \ . \la{216}
\ee
The resulting $Z_1$ has the form \rf{101} where 
 the  one-particle partition function  is thus equal to the character  of    spin 1 doubleton representation in \rf{a10} 
  (cf. \rf{212})
\be
\Z_{1}  = \mZ(\{1,0\}_c) = \sum_{n=0}^{\infty} \dd_{n}\   e^{-\b \omega_n } = \frac{2q^2\, (3-q)}{(1-q)^{3}} \ .  \la{217}
\ee
Let   us now consider $N$ complex Maxwell   vectors and impose  the singlet constraint. 
One way to do this is to start with  $U(N+1)$   YM   theory  and split  the $U(N+1)$ field 
into the $U(N)$ one $\AA_m$,  $N$ complex vectors $A_m$  in the fundamental of $U(N)$  and a  singlet.
Then the YM action can be written as in  \rf{71} 
\be 
L=  { 1 \ov 2}   F^*_{r\, mn}   F^{mn} _r +   { 1 \ov 4 g^2}  \tr \, \FF_{mn} \FF^{mn}  \ , 
\ \ \ \ \ \quad     F^r_{mn} = \cD_m A^r_n -  \cD_n A^r_m  \ ,         \la{7111} \ee
where $\cD_m = \cD_m (\AA)$, $r=1, ..., N$. We rescaled $A_m$ by $g$   and ignored  the decoupled singlet. 
Taking the limit $g \to 0$ (for fixed $N$)  
 understood as localizing the 
path integral over $\AA_m$ on $\FF_{mn}=0$ or pure gauge configurations 
we get then 
  the direct analog of \rf{71},\rf{177}, i.e. the following generalization of \rf{a0}
  \be L_1 =    \cD_0 A^*_{r\, i}  \cD_0 A^i_r  +  { {1 \ov 2}} F^*_{r\, ij} F^{ij}_r   
   \ ,  \la{a00} \ee 
where $\cD_0$  contains  $\A_0$ in \rf{177}  (we   ignored the trivial decoupled contribution of $A_0$). 
Integrating over $A^r_i$ and $U$  we then  get   again the relations \rf{55},\rf{5b},\rf{5a}  where now 
$\dd_n$ and $\om_n$ are given by \rf{216}. 
The  remaining derivation of the  large $N$ limit of  the 
singlet partition function is then literally 
 the same as above  in the  scalar case. 

As   a result, we get exactly the same   expressions for the singlet-sector partition  function 
 as in \rf{17},\rf{177}    with $\Z_0 $ replaced by $\Z_1$, i.e.\foot{Note that the resulting canonical 
 partition function in 
 \rf{178} is different from the canonical single-trace   partition function   of  one-loop 
$SU(N)$   YM theory  on $S^1 \times S^3$  \ci{Sundborg:1999ue,Polyakov:2001af,Aharony:2003sx} 
counting single-trace operators  built out of any number of fields, not just two. It is 
 given at large $N$  by $\Z_{\rm YM} = - \sum_{k=1}^\infty  {\vp(k) \ov k} \ln [ 1 - \Z_1 ( k \b) ]$,  
 where $\vp$ is Euler's  totient function.} 
 \ba   & 
  \hat \Z_{\rm U(N)} (\b)    =   \big[ \zz_1 (\b)\big]^2  = \Big[{ 2 q^2 (3- q) \ov (1-q)^3} \Big]^2   \  ,   \la{v17} \\
  &
  \hat \Z_{\rm O(N)}    (\b) =    \ha \big[ \zz_1 (\b)\big]^2   + \ha \Z_1 (2\b)  =     \ha  \Big[{ 2 q^2 (3- q) \ov (1-q)^3} \Big]^2    + \ha  { 2 q^4 (3- q^2) \ov (1-q^2)^3}      \  .  \la{v77}
\end{align} 
These   expressions are indeed consistent with the AdS/CFT, i.e. with counting of conserved currents or 
counting of states in the  dual type C theory  in \ads  as follows from the comparison 
with  the  identities   for the corresponding  characters  in \rf{a7} and \rf{a89}. 

\renewcommand{\theequation}{4.\arabic{equation}}
\setcounter{equation}{0}
\section{Casimir energy} 

The    Casimir energy is determined by the same spectrum as  the one-particle  partition function 
$\Z$  in \rf{101},\rf{2222}   and   they may be directly  related as  
\be
E_c= \ha \,  \sum_n  \dd_n\, {\om_n}  = \ha \,\zeta_E (-1) \ , \ \ \qquad \  \zeta_E (z) =\sum_n  {\dd_n \ov {\om^z_n}}   =  {1\ov \Gamma(z) } \int^\infty _0 d \beta \, \beta^{z-1} \, \Z(e^{-\b}) \ . \la{2281}  
\ee
One   can  show  that the Casimir energy { vanishes}  if the partition function $\Z$   obeys the property 
 $\Z(q) = \Z(q^{-1} )$  \cite{Giombi:2014yra}. 
 This   property is true for the square of the scalar partition function in \rf{17}  (cf. \rf{210})
 but is not   valid  for the second term in \rf{177}  in the  $O(N)$ case.
  Since  the Casimir energy is given by the  linear in $\b$ part of the log of the total partition function in \rf{101}
    it then follows that $E_c$  for the spectrum 
 corresponding to the singlet partition function \rf{177}  is  the same as for one
  real   4d   scalar, i.e. in the scalar or type A  case of AdS/CFT we have   
 \be \la{81}
{\rm scalar}:\qquad   (\hat E_c)_{\rm U(N)} =  0 \ , \ \ \ \ \ \ \ \ \  \ \ \   (\hat E_c)_{\rm O(N)} =  E_{c\, 0} =  E_{c} (\{0,0\}) 
    =\te  { 1 \ov 240}  \ . 
 \ee 
 In the vector case \rf{v17} there is no $q\to q^{-1}$ invariance already in the $U(N)$ case.
 Since  the combination 
 $[\Z_1(\b)]^2 - 2 \Z_1(\b) = - {4 (3-q) (1- 3q) q^2 \ov (1-q)^6}$ is symmetric  under $q \to q^{-1}$  
   in this case the singlet-sector Casimir energy is same as of 2 real  Maxwell 
 4d vectors. 
 In the $O(N)$ case  the contribution  of the first term in    \rf{v77}
   is  half of the   $U(N)$   case one  and the second term  gives   the same 
$E_{c\, 1} $   contribution.  Thus we find 
 \be \la{81a} 
 {\rm vector}:\qquad    (\hat E_c)_{\rm U(N)} = (\hat E_c)_{\rm O(N)} =  2 E_{c\, 1} = 2 E_c(\{1,0\}_c)=\te 2 \times { 11 \ov 120}  \ . 
 \ee 
The  relation between  the  characters   or one-particle partition functions in \rf{a20}.\rf{a21}
then implies the   corresponding  relations for  the  total Casimir energy of the \ads   theory:  
in non-minimal  and minimal type C theory  we then get, respectively,  
\ba
 &
  2\, E_c(4; 0, 0)+   E_c(4; 1,0)_{c}  
  +    2  \sum_{s=2}^{\infty} E_c (2+s; {\ts\frac{s}{2}, \frac{s}{2}})  +   \sum_{s=2}^{\infty} 
E_c(2+s; {\ts\frac{s+2}{2}, \frac{s-2}{2}})_{c}  =    2   E_{c\, 1}   \ , \la{a200}   \\
& 
 2\, E_c(4; 0, 0)+\sum_{s=2}^{\infty}  E_c (2+s; {\ts\frac{s}{2}, \frac{s}{2}})
    +  \sum_{s=2,4,\dots}^{\infty}  E_c(2+s; {\ts\frac{s+2}{2}, \frac{s-2}{2}})_{c}  =  2 E_{c\, 1}  \ . 
   \la{a201} \end{align}
Note that  here one  does not  need  to worry about 
  regularization  of the sums  over  spins 
 provided one  uses the $\zeta$-function  prescription to define $E_c$  as in \rf{2281}. Namely, one is first to compute
  the 
 sum  over  spins of  all  partition functions for finite $q$ or  the  sum of their Mellin  transforms $\zeta_E(z)$  
  and  then to continue  the result  to the required  $z=-1$  value.\foot{This procedure is equivalent to using an exponential cutoff
   $\exp [ - \ep ( s + \ha) ]$  in the sum over $s$ and dropping all terms that are singular in the $\ep\to 0$ limit, see \ci{Giombi:2014yra}.}
 In view of \rf{a7},\rf{a20} and \rf{81a}   the relation  \rf{a200} 
 expresses   the equality of $E_c$ computed for the product  of two  spin 1 doubleton representations 
   to twice  $E_c$ for the single   doubleton  representation, i.e. 
   \be \la{aaa}
   E_c \big( \{1, 0\}_c \otimes   \{1, 0\}_c\big) = 2    E_c \big(  \{1, 0\}_c\big)  \ . \ee
Remarkably, the   relations \rf{a200} or \rf{aaa} and \rf{a201}   are  true also for the 
 boundary theory conformal  anomaly coefficients 
$a$ and $c$ \ci{Beccaria:2014xda} (these correspond to  partition functions for  $S^4$ or Ricci flat space instead of  $S^1 \times S^3$). 

To  interpret \rf{aaa}  from the point of view of AdS/CFT  we observe  that   the full  boundary  CFT contribution to the Casimir energy 
is simply  proportional to  $N$ 
 (cf. \rf{178}).  This  should correspond to  the classical  type C theory  contribution in \adss.
Then the duality    requires  the  one-loop   bulk  contribution to $E_c$ to  vanish, but it does not  according to \rf{a200},\rf{a201}. 
To reconcile this with AdS/CFT  duality 
we  follow the suggestion made  in the real scalar and fermion cases 
 \ci{Giombi:2013fka,Giombi:2014yra}   and  conjecture  that the coefficient in  front of the classical  bulk theory 
 action  should  be  shifted   by an integer  from its naive   value $N$.
 Namely,  let us assume  that the classical actions  of   the non-minimal type C theory  dual to 
  complex $U(N)$  Maxwell theory  and  the minimal type C theory  dual to 
  real  $O(N)$  Maxwell theory   have the form
  \be  S_{\rm non-min \ U(N)} =  (2 N-2)  \big(S_0 + ...\big)  \ , \ \ \ \ \ \ \ \  S_{\rm  min \ O(N)} = (N-2) \, S_0 \ . \la{bb} \ee
  Here $S_0$ stands  for the common sector of the two type C  theories (cf. \rf{a7},\rf{a89})
   Then the factor of  two   difference  between the leading  large $N$ terms  in the two actions 
   will   be consistent with the fact that 
   the  boundary theory 
   Casimir energy of $N$ complex fields is the same as  that  of the $2N$ real fields. 
    The equal  negative  subleading terms  will be required to cancel 
    the  equal  one-loop corrections  \rf{a200},\rf{a201}  of the  quantum fluctuations of the bulk fields, 
    in agreement with the absence of the subleading in $N$  term in $E_c$   in the  boundary theory. 
    
  It is interesting to note  that  the structure of the two  actions in \rf{bb} is consistent also with the  fact that 
  $U(N)$   theory should be equivalent to $O(2N)$  one  as far as  ``non-singlet'' 
  properties like  Casimir energy  or partition function on  a space with trivial holonomy (e.g., $S^4$) 
    are concerned: one complex field   should be  the same as 2 real ones. The reason for the triviality of the 
    bulk  action  in the $U(1)$  ($N=1$)  case  implied by \rf{bb}  remains to be clarified further. 
 
 The above  discussion has a straightforward generalization to the case when  the boundary 
 theory is  represented  by a combination of vectors, fermions and scalars and, in particular, to the supersymmetric case. 
  For example, the   relation \rf{aaa}  holds  also if one  replaces   the spin 1  doubleton by   a  combination 
$\{1, 0\}_c + n_{1\ov 2} \{\ha , 0\}_c+ n_0 \{0, 0\}$  which represents  a superdoubleton for 
special choices of $n_{1\ov 2}$ and $n_0$ \ci{Beccaria:2014xda}. 
  In the case of $\N=4$   superdoubleton ($n_{1\ov 2}=4, \ n_0=6$)
one gets the  duality between the singlet sector of the theory  of  $N$  copies of $\N=4$   supersymmetric Maxwell  multiplet 
and a  special  $\N=4$ supersymmetric higher spin theory in \ads   generalizing 
maximally supersymmetric   5d  gauged supergravity   \ci{Gunaydin:1998sw,Sezgin:2001zs,Sezgin:2002rt}.

\section*{Acknowledgments}
We  thank K. Alkalaev, R. Metsaev   and  E. Skvortsov
  for   important    discussions. 
This  work   
was  supported   by  the  Russian Science Foundation grant 14-42-00047 associated  with  Lebedev Institute.
The  work of A.A.T.  was  also partially supported by the ERC Advanced grant No.290456.

 
 \newpage 
 \baselineskip 8pt
 


\end{document}

\end{document}